\documentclass[10pt, twocolumn, amsmath,amssymb,eqsecnum,prd,aps,superscriptaddress, nofootinbib]{revtex4-2}
\usepackage{newunicodechar}
\newunicodechar{−}{\ensuremath{-}}

\usepackage{silence}
\usepackage{geometry}
\usepackage{multirow}
\usepackage{booktabs}
\usepackage{tabularx}
\usepackage{adjustbox}
\DeclareUnicodeCharacter{03C3}{-}
\DeclareUnicodeCharacter{03C9}{-}
\usepackage{amsmath,amssymb}
\usepackage{amsthm}
\usepackage{mathtools}
\usepackage{appendix}
\usepackage{comment}
\usepackage{graphicx}
\usepackage{grffile}
\usepackage{tikz}
\usepackage{physics}
\usepackage{tikz-feynman}
\usepackage{mathrsfs}
\usepackage{bm}
\usepackage{url}
\usepackage[table]{xcolor}
\usepackage{orcidlink}
\usepackage{xcolor}
\usepackage{soul}
\usepackage{algorithm}
\usepackage{algorithmic}
\usepackage{natbib}
\usepackage{siunitx}
\usepackage{acronym}
\usepackage{xspace}
\usepackage{subfloat}
\usepackage[normalem]{ulem}
\usepackage{amssymb}
\usepackage{amsmath}
\usepackage{graphicx}
\usepackage{dcolumn}
\usepackage{hyperref}
\usepackage{float} 
\usepackage{multirow}
\usepackage{amsfonts,wasysym,epsfig,verbatim,subfigure,bm,mathrsfs,lipsum}
\usepackage{bm}
\usepackage{soul}
\usepackage{pifont}
\usepackage{multirow}
\usepackage{natbib}
\usepackage{ragged2e}

\definecolor{MONZA}{HTML}{CF000F}
\definecolor{DARKBLUE}{HTML}{00008b}
\definecolor{DARKMAGENTA}{HTML}{8b008b}
\definecolor{DARKCYAN}{HTML}{00cfc0}
\definecolor{brightpink}{rgb}{1.0, 0.0, 0.5}

\def\nn{\nonumber}

\newcommand{\orcidicon}[1]{\href{https://orcid.org/#1}{\includegraphics[height=\fontcharht\font`\B]{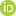}}}

\begin{document}


\title{First Constraints on the Ellipticities of Self-Interacting Fermionic Dark Matter–Admixed Neutron Stars from Continuous Gravitational-Wave Searches}

\author{Premachand Mahapatra~\orcidlink{0000-0002-3762-8147}}
\email{p20210039@goa.bits-pilani.ac.in}
\affiliation{Department of Physics, Birla Institute of Technology and Science-Pilani, K. K. Birla Goa Campus, NH-17B, Zuarinagar, Sancoale, Goa- 403726, India}

\author{Andrew L. Miller~\orcidlink{0000-0002-4890-7627}}\email{andrew.miller.ligo@ucas.ac.cn}
\affiliation{International Center for Theoretical Physics Asia-Pacific (ICTP-AP), University of Chinese Academy of Sciences, Beijing 100190, China}
\affiliation{Taiji Laboratory for Gravitational Wave Universe, University of Chinese Academy of Sciences, 100049 Beijing, China}

\author{Prasanta Kumar Das~\orcidlink{0000-0002-2520-7126}}
\email{pdas@goa.bits-pilani.ac.in}
\affiliation{Department of Physics, Birla Institute of Technology and Science-Pilani, K. K. Birla Goa Campus, NH-17B, Zuarinagar, Sancoale, Goa- 403726, India}

\date{\today}

\begin{abstract}

We investigate continuous gravitational-wave (CW) emission from rapidly rotating, non-axisymmetric, isolated neutron stars admixed with self-interacting fermionic dark matter (DM) and hosting DM-induced equatorial deformations (``dark mountains''). In particular, we develop a formalism that describes how DM accumulation inside the star changes its structure, how dark mountains arise from an anisotropic distribution of DM inside it, and how the star's moment of inertia and thus the amplitude of its GW emission is increased compared to that of an ordinary neutron star. Moreover, using results from all-sky searches for CWs from non-axisymmetric neutron stars performed with LIGO O3 data, we place the first constraints on the DM-induced ellipticities $\varepsilon$ of DM-admixed neutron stars across the full GW frequency range analyzed by LIGO and for a range of self-interaction strengths. With the same data, we also exclude portions of the DM-mass/self-interaction coupling strength parameter space that would have produced detectable GW signals in LIGO O3 data. We rule out at best (at worst) couplings $g\gtrsim10^{-5.5}$ ($g\gtrsim 10^{-4}$) for DM-admixed neutron stars with ellipticities $\varepsilon=10^{-7}$ ($\varepsilon=10^{-9}$) at distances $d=1$ ($d=10$) kpc away for DM masses of $m_\chi\in[0.1,10]$ GeV. Furthermore, we show that even larger regions of this parameter space will become accessible to next-generation detectors, such as Einstein Telescope and Cosmic Explorer, with exclusions as strong as $g\gtrsim10^{-6}$ for neutron stars located $d=10$ kpc away for $\varepsilon=10^{-7}$. Our results demonstrate that searches for CWs naturally provide a direct probe of dark mountains sustained by DM-admixed neutron stars. \\
%

\end{abstract}

\maketitle
\acresetall

\acrodef{SNR}{signal-to-noise ratio}
\acrodef{PSD}{power spectral density}
\acrodef{PBH}{primordial black hole}
\acrodef{NS}{neutron star}
\acrodef{WD}{white dwarf}
\acrodef{GW}{gravitational wave}
\acrodef{NFW}{Navarro--Frenk--White}
\acrodef{LVK}{LIGO--Virgo--KAGRA}
\acrodef{CW}{Continuous Gravitational Waves}
\acrodef{DM}{Dark matter}

\section{Introduction}

The direct detection of gravitational waves (GW) from merging black holes and neutron stars~\cite{LIGOScientific:2016aoc, LIGOScientific:2017vwq} has transformed our understanding of compact objects and marked the beginning of a new era in relativistic astrophysics. The global network of ground-based interferometers -- LIGO, Virgo, and KAGRA (LVK) -- has subsequently established GW astronomy through the detection of $\mathcal{O}(10^{2})$ compact binary coalescence events over multiple observing runs~\cite{LIGOScientific:2026wfs}. These observations consist entirely of transient, chirping signals produced by merging binary black holes and neutron stars. In contrast, an equally compelling and astrophysically rich class of signals -- \emph{continuous gravitational waves} (CWs) -- has not yet been detected \cite{Riles:2017evm, Piccinni:2022vsd, Riles:2022wwz}. CWs are defined as persistent, quasi-monochromatic signals, whose amplitudes and frequencies change orders of magnitude more slowly than those of transient sources. Such signals can originate from a variety of astrophysical and particle-physics sources, including non-axisymmetric rotating neutron stars \cite{Jones:2001yg, Woan:2018tey}, ultralight dark matter (DM) clouds around spinning black holes \cite{Brito:2015oca, Brito:2017zvb}, ultralight particle DM that couples directly to ground-based interferometers \cite{Pierce:2018xmy, Grote:2019uvn, Michimura:2020vxn, Armaleo:2020efr, Hall:2022zvi}, and inspiraling planetary-mass primordial black holes (PBHs) \cite{Miller:2020kmv, Miller:2024rca}. While many efforts to find deformed neutron stars~\cite{KAGRA:2022dwb, Steltner:2023cfk, Covas:2024nzs, LIGOScientific:2025kei, LIGOScientific:2025ouy, McGloughlin:2025eso, McGloughlin:2025iyx,Ming:2025ehy,Clark:2025lai,Dergachev:2025ead,Covas:2026raf}, particle dark matter \cite{Guo:2019ker, Nagano:2019rbw, Miller:2020vsl, Vermeulen:2021epa, Miller:2022wxu, Manita:2023mnc, Gottel:2024cfj, KAGRA:2024ipf, LIGOScientific:2025csr}, boson clouds \cite{DAntonio:2018sff, LIGOScientific:2021rnv, LIGOScientific:2025ttj} and PBHs \cite{Miller:2021knj, Miller:2024fpo, LIGOScientific:2025vwc} have been made, such objects have all eluded detection.

{
In the absence of a detection, upper limits are set on the minimum detectable strain amplitude visible at the detector. These upper limits can be directly related to the neutron-star properties because the CW amplitude depends on the equatorial ellipticity $\varepsilon$ and the $z$-component of the principal moment of inertia $I_{\rm zz}$. Consequently, CW observations constrain the allowed combinations of $(\varepsilon, I_{\rm zz})$ for potential sources, providing a direct link between observational sensitivity and the internal structure of neutron stars \cite{Ushomirsky:2000ax,Cutler:2002nw, Soldateschi:2021hfk, Biryukov:2025zga,Jones:2001yg, Gao:2020zcd,Bildsten:1998ey, Melatos:2007zz,Andersson:1997xt, Dong:2025roh}. In a similar way, we can also use CW upper limits to constrain the presence of non-standard exotic matter (e.g. dark matter) inside the neutron star, which can enhance quadrupolar deformations and hence GW emission~\cite{Bhattacharya:2024pmp, Miller:2025evt, Pagliaro:2025qrr}. At the same time, though, recent theoretical constraints have begun to challenge the abundance of such exotic compact objects, with some analyses disfavoring their existence~\cite{Prabhu:2024lzl}. This tension motivates continued CW searches: as we will show, tighter limits on strain directly constrain non-standard neutron-star models and exotic-matter scenarios. 


A well-motivated model for exotic neutron stars involves the presence of DM inside the star. DM can be captured in the core during core-collapse supernova~\cite{ Guver:2012ba, Xiang:2013xwa, DelPopolo:2020hel} or accreted throughout the star's lifetimes~\cite{Bertone:2007ae, Ilie:2020nzp, Koehn:2024gal}, leading to the formation of a stable two-fluid configuration of baryonic matter (BM) and DM within the star~\cite{Leung:2011zz,  Petraki:2013wwa, Kouvaris:2015rea, Mukhopadhyay:2016dsg, Ellis:2018bkr}. Despite extensive work on how such an admixture of DM alters the macroscopic properties of neutron stars, such as internal density, mass-radius relations, tidal deformabilities, cooling behaviour, and stability~\cite { Sagun:2021oml, Bhattacharjee:2024pis, Bhat:2019tnz, Das:2020ecp, Haskell:2022pqt, Lu:2022oys, Grippa:2024ach}, the implications of such admixtures for CW emission remain underexplored\footnote{One study evaluated the impact of self-interacting DM on $r$-mode oscillations in neutron stars, which showed that DM-induced modifications to shear and bulk viscosities can substantially shift the $r$-mode instability, with direct implications for GW emission~\cite{Shirke:2023ktu}.}.

 In particular, a class of self-interacting fermionic DM of mass $m_\chi$, mediated by Yukawa-type couplings $g$ ~\cite{Janish:2019nkk, Mahapatra:2024ywx, Liu:2025cwy}, can accumulate inside neutron stars~\cite{Petraki:2013wwa}, while remaining consistent with astrophysical bounds from the Bullet-Cluster~\cite{Randall:2007ph, Tulin:2017ara} and direct-detection experiments~\cite{XENON:2018voc, DarkSide:2018bpj}, provided that the self-interaction cross-section satisfies the observationally-allowed constraints of $\sigma/m_\chi \sim 0.1$ -- $10 \,\text{cm}^2/\text{g}$)~\cite{Markevitch:2003at, Kaplinghat:2015aga}, where $\sigma$ is the DM cross section. Older theoretical studies on this class of DM-admixed neutron stars primarily focused on heavy mediators at MeV masses with relatively large couplings $g\sim10^{-2}$--$10^{-1}$~\cite{Bramante:2013nma}, while more recent ones explored relic-density-motivated self-interacting DM scenarios for $m_\chi \sim 0.1-30$ GeV~\cite{Guha:2024pnn}. 

In contrast to the aforementioned theoretical works, two other studies addressed the extent to which such DM-admixed neutron stars could be constrained with real data. One relied on pulsar measurements and the first binary neutron star merger GW170817 to constrain $g \sim 0.01$ -- $0.1$ for mediator masses in the MeV range \cite{Mariani:2023wtv}. Using the same observational inputs, another explored mediator masses in the keV range and determined that significantly smaller self-interaction couplings in the range $g\sim10^{-6}$--$10^{-4}$ are allowed, which also permits the possibility of core-halo transitions in DM-admixed neutron stars~\cite{Mahapatra:2024ywx}.

In contrast to previous works, here we study, for the first time, the impact of self-interacting fermionic DM on the CW emission from rotating neutron stars in the $(g, m_\chi)$ parameter space. We consider $m_\chi \in [0.1,10]$~GeV spanning the sub-GeV to multi-GeV regime, and $g \in [10^{-6},10^{-3.5}]$, consistent with astrophysical limits on $\sigma/m_\chi$, and we develop a way to set constraints on this DM parameter space using null results from CW searches for neutron stars. Using a two-fluid DM-admixed neutron star model, we calculate the expected CW strain amplitudes as a function of $m_{\chi}$ and $g$, and compare them with upper limits derived from an all-sky search for isolated neutron stars using data from the third observing run (O3)\footnote{Shortly before we finished our study, the LVK released upper limits on strain amplitude from an all-sky search for neutron stars using data from the first part of the fourth observing run (O4a) \cite{LIGOScientific:2025ouy}. However, the best constraint on the CW amplitude is only about 10\% better than what we used throughout this work, so our results would not qualitatively change.}. With these results, we can rule out regions of $m_{\chi}-g$ parameter space that could source DM-admixed neutron stars, and identify those that will be excluded with next-generation detectors, such as the Einstein Telescope (ET) \cite{ET:2019dnz, Branchesi:2023mws, ET:2025xjr} and Cosmic Explorer (CE) \cite{Reitze:2019iox, Evans:2023euw, DiGiovanni:2025rhy}.

The paper is organized as follows. In Sec.~\ref{sec:formalism}, we review the theoretical framework of CW emission, and the two-fluid DM-admixed neutron star model. In Sec.~\ref{sec:dm_deformation}, we derive how dark mountains on an isolated neutron star can be sourced through an anisotropic DM distribution inside the star. In Sec.~\ref{sec:methodology}, we outline our procedure to connect theoretical predictions of CWs from DM-admixed neutron stars with upper limits on strain amplitude from the LVK all-sky search for CWs using data from the third observing run. In Sec.~\ref{sec:results}, we show the results of our work, including how the CW amplitude changes as a function of GW frequency, the maximum sustainable ellipticity permitted by null results from O3, and exclusion contour plots in the ($g, m_\chi$) DM parameter space. In Sec.~\ref{sec:implications}, we demonstrate how future GW detectors can be used to constrain DM-admixed neutron stars. Finally, in Sec.~\ref{sec:conclusion}, we summarize our findings and their implications for future GW observations.

}

\section{Formalism}
\label{sec:formalism}

\subsection{Continuous Gravitational Waves}

Considering a neutron star as an oblate spheroid, its mass quadrupole moment can be characterized by an equatorial ellipticity $\varepsilon$, i.e. a ``mountain" \cite{Maggiore:2007ulw, Riles:2022wwz, Piccinni:2022vsd,  Wette:2023dom, Jones:2024npg}:

\begin{equation}
    \varepsilon \equiv \frac{|I_{\rm xx} - I_{\rm yy}|}{I_{\rm zz}} 
\end{equation}
where $I_{\rm xx}, I_{\rm yy}$ and $I_{\rm zz}$ are the principal moments of inertia about each axis, given by~\cite{Maggiore:2007ulw} 

\begin{align}
I_{\rm xx} &= \int_V \rho(\mathbf{x}) \left(y^2 + z^2\right)\, dV \label{Ixx} \\
I_{\rm yy} &= \int_V \rho(\mathbf{x}) \left(z^2 + x^2\right)\, dV \label{Iyy} \\
I_{\rm zz} &= \int_V \rho(\mathbf{x}) \left(x^2 + y^2\right)\, dV \label{Izz}
\end{align}
$\rho(\mathbf{x})$ is the position-dependent mass density of the star, and $V$ is the stellar volume.

In terms of the GW frequency $f_{\rm GW}$ and $\varepsilon$, we can write the CW amplitude $h_0$ for a star at a distance $d$ away that is spinning about the $z$-axis \cite{Maggiore:2007ulw, Riles:2022wwz, Wette:2023dom, Jones:2024npg} as:

\begin{align}
h_0 &= \frac{4\pi^2 G}{c^4}\,
      \frac{\varepsilon I_{\rm zz} f_{\rm GW}^2}{d}
\nonumber\\
&\simeq 1.06 \times 10^{-25}
\left(\frac{\varepsilon}{10^{-9}}\right)
\left(\frac{I_{\rm zz}}{I_0}\right)
\left(\frac{10\,{\rm kpc}}{d}\right)
\left(\frac{f_{\rm GW}}{1\,{\rm kHz}}\right)^2 ,
\label{eq:h0}
\end{align}
where $ I_0 = 1 \times 10^{38} \, \text{kg m}^2 $ is the nominal moment of inertia value used in \cite{Maggiore:2007ulw}. 



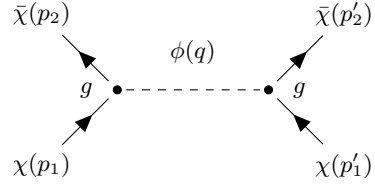
\begin{figure}[htbp]
   \centering
    \begin{tikzpicture}
        \begin{feynman}
            \vertex (v1) at (-1,0) {};  
            \vertex (v2) at (1,0) {};   
            
            \vertex (a1) at (-2,1) {$  \bar{\chi}(p_2)$};
            \vertex (b1) at (-2,-1) {$\chi (p_1)$};
            \vertex (a2) at (2,1) {$\bar{\chi}(p'_2)$};
            \vertex (b2) at (2,-1) {$\chi(p'_1)$};
            \vertex (phi) at (0,0.5) {$\phi(q)$}; 
            
            \diagram* {
                (b1) -- [fermion] (v1) -- [fermion] (a1),  
                (b2) -- [fermion] (v2) -- [fermion] (a2),  
                (v1) -- [scalar, dashed] (v2), 
            };
            
            \filldraw (v1) circle (1.5pt);
            \filldraw (v2) circle (1.5pt);
            
            \node[left=6pt] at (v1) {$g$};
            \node[right=6pt] at (v2) {$g$};
        \end{feynman}
    \end{tikzpicture}
    \caption{$\chi \bar{\chi}$ scattering mediated by a scalar field $\phi$ in the $s$-channel with coupling $g$.}
    \label{fig:feynmandiagram_SIDM}
\end{figure}

{
Equating the power emitted due to GW emission with the loss of rotational kinetic energy 
we can derive an expression for how the rotational frequency of a neutron star decreases over time, known as the spin-down $\dot{f}_{\rm GW}$~\cite{Maggiore:2007ulw, Riles:2022wwz}

\begin{align}
\dot{f}_{\rm GW} &=
-\,\frac{32\pi^4 G}{5c^5}\,
\varepsilon^2 I_{\rm zz} f_{\rm GW}^5
\nonumber\\[3pt]
&\simeq -\,1.7\times10^{-9}\,{\rm Hz\,s^{-1}}
\left(\frac{\varepsilon}{10^{-9}}\right)^2
\left(\frac{I_{\rm zz}}{I_0}\right)\left(\frac{f_{\rm GW}}{1\,{\rm kHz}}\right)^5 .
\label{eq:fdot}
\end{align}
Eq. \eqref{eq:h0} and Eq. \eqref{eq:fdot} demonstrate that $h_0$ and $\dot{f}_{\rm GW}$ depend on $I_{\rm zz}$, $\varepsilon$ and $f_{\rm GW}$. These quantities in turn are determined by an internal mass distribution that is a function of the composition of stellar matter.

Therefore, any additional matter that alters the star's mass distribution can modify its quadrupole moment and rotation, leading to corresponding changes in the CW emission. In particular, the presence of DM can affect $I_{\rm zz}$ and $\varepsilon$, which will be quantified for DM-admixed neutron stars in the following subsection.

\subsection{DM-Admixed Neutron Stars }
\label{subsec:dmns}


DM-admixed neutron stars consist of both baryonic matter and DM, and have a total mass density of

\begin{align}
    \rho_{\text{tot}} = \rho_{\rm B} + \rho_{\chi},
\end{align}
where $\rho_{\rm B}$ denotes the baryonic matter density and $\rho_{\chi}$ is the fermionic DM density. 

To calculate $\rho_{\rm B}$, we use the relativistic mean-field framework that provides a well-established description of dense nuclear matter in neutron stars~\cite{Dutra:2014qga, Zhu:2023ijx}






\begin{align}
    \rho_{\rm B} = \rho_{\rm RMF} = 2.6 \times 10^{18} \, \text{kg m}^{-3} \sim \mathcal{O}(10^{18}) \, \text{kg m}^{-3},
    \label{eq:matterdensityRMF}
\end{align}
which is consistent with typical neutron-star interior densities. The corresponding $z$-component of the moment of inertia, assuming the density is constant and canonical radius of $R_{\rm NS}= 10 \, \mathrm{km}$, is 


\begin{align}
I_{\rm zz}\big|_{\rho_{\rm B}}
&= 4.13 \times 10^{38}\,{\rm kg\,m^2}
   \;\approx\; \mathcal{O}(10^{38})\,{\rm kg\,m^2} .
\label{izzrmfns}
\end{align}
Because $\rho_{\textrm B}$ is fixed by the underlying nuclear model and does not vary with the parameters explored in this study, it sets the baseline structure of the star. 
{In addition to baryonic matter, we assume that gravitationally-bound self-interacting fermionic DM resides in the neutron star. 
We model the self-interaction of this DM $\chi$ with a Lorentz scalar field $\phi$, with the Lagrangian 
\begin{equation}
    {\mathcal L} = g \overline{\chi} \chi \phi.
\end{equation}} 
This self-interacting DM will contribute to the process $\chi \overline{\chi} \to \chi \overline{\chi}$ mediated by $\phi$, as shown in the Feynman diagram in Fig. \ref{fig:feynmandiagram_SIDM}.}
{The scale of this interaction is set by the mediator mass $m_\phi$, which, if small enough, sets the range of interaction between DM particles to be sufficiently long (but finite).

Not only does the DM mass itself contribute to the overall density of the star, but the self-interaction does as well, which can be calculated as follows.
In the non-relativistic regime, the potential energy between two DM particles  of charges $g$ is

\begin{equation}
    V_{ij}(r_{ij}) = \frac{g^2}{4 \pi} \frac{e^{- m_\phi r_{ij}}}{r_{ij}},
\end{equation}
which is of Yukawa type and repulsive in nature. Note that as $m_\phi \to 0$, the potential becomes a repulsive Coulomb type with an infinite range. 

For a system of $N$ self-interacting particles (i.e. DM fermions) in volume $\Omega$, the total Yukawa potential energy is estimated as \cite{Kouvaris:2015rea, Mukhopadhyay:2016dsg}
\begin{equation}
    E_\Omega =  \frac{1}{2} \sum_{i \neq j} V_{ij} = \frac{1}{2} \frac{n^2 g^2}{4 \pi} \int \int  \frac{e^{- m_\phi r_{ij}}}{r_{ij}} d\Omega_i d\Omega_j = \frac{n^2 g^2 \Omega}{2 m_\phi^2},
\end{equation} 
where $n$ is the DM number density. Assuming that the DM fermions form a degenerate Fermi gas inside the neutron star, the number density can be calculated as 
\begin{equation}
n = \frac{N}{V} =  \frac{2 s + 1}{(2 \pi)^3} \left(\frac{4}{3} \pi k_F^3\right) = \frac{1}{3 \pi^2} k_F^3,    
\end{equation} where $s = 1/2$ for DM fermions.

For simplicity, writing $x \equiv k_F/m_\chi$, where $k_F = (3 \pi^2 n)^{1/3}$ is the Fermi momentum in natural units ($\hbar = c = 1$), we finally find the contribution of the self-interaction to the DM energy density:
\begin{equation}
    \rho_Y = \frac{E_\Omega}{\Omega} = \frac{g^2}{2 m_\phi^2 } \frac{k_F^6}{(3 \pi^2)^2} = 
\frac{1}{(3 \pi^2)^2 } \frac{g^2 x^6 m_\chi^2}{ 2 m_\phi^2}.
\end{equation}
}
The total DM energy density also consists of a kinetic energy contribution and can be written as \cite{Kouvaris:2015rea, Mukhopadhyay:2016dsg}

\begin{equation}
\label{eq:ferm}
    \begin{split}
       \rho_{\chi} =& \rho_{\rm kin}(x) + \rho_{Y}(x) =  \Bigl\{ \frac{m_\chi^4}{8 \pi^2} \Bigl( x \sqrt{1+x^2} (1 + 2 x^2) \\
        &- \ln\left(x + \sqrt{1+x^2} \right) \Bigr) \Bigr\} + \frac{g^2 x^6 m_\chi^6}{2 (3 \pi^2)^2 m_\phi^2} \\
     \end{split}
\end{equation}

Rather than focusing on a single benchmark parameter space of DM, we systematically explore a discrete grid of $m_{\chi}$ and $g$ values: 
\begin{align}
m_\chi &\in [0.1,\;10.0]\;\mathrm{GeV}, \\
g &\in [10^{-6},\;10^{-3.5}],
\end{align}
and fix the mediator mass of $m_\phi = 1 \,\mathrm{keV}$ to satisfy the observationally-allowed self-interaction cross-sections ($\sigma/m_\chi \sim 0.1–10 \,\text{cm}^2/\text{g}$) for self-interacting DM ~\cite{Markevitch:2003at, Kaplinghat:2015aga} (see App. A for a derivation of $\sigma/m_\chi$).

Our choices are motivated by previous studies of self-interacting fermionic asymmetric DM (ADM) in neutron stars~\cite{Nelson:2018xtr, Rutherford:2024uix}. In particular, Nelson \textit{et al.}~\cite{Nelson:2018xtr} studied fermionic ADM particles with masses of order $\mathcal{O}(10^2)\, \mathrm{MeV} \sim \mathcal{O}(10^{-1})\, \mathrm{GeV}$ interacting through a light vector mediator in the eV--keV mass range, which can induce changes to neutron-star macroscopic properties, such as mass and radius. More recently, Rutherford \textit{et al.} \cite{Rutherford:2024uix} investigated fermionic ADM cores in neutron stars within a Bayesian framework and considered the prior ranges, which are almost identical to their corresponding 2-D posterior distribution, of
$m_\chi \in [10^{-5},10^{6}] \,\mathrm{GeV}$ and
$g \in [10^{-8},1]$.


The chosen mass interval spans the transition from the sub-GeV regime explored in~\cite{Nelson:2018xtr} to the multi-GeV regime considered in self-interacting ADM cases. At the same time, the adopted coupling range constitutes a physically motivated subset of the broader parameter space investigated by~\cite{Rutherford:2024uix}.


\begin{figure}[htbp]
    \centering
    \includegraphics[width=0.95\linewidth]{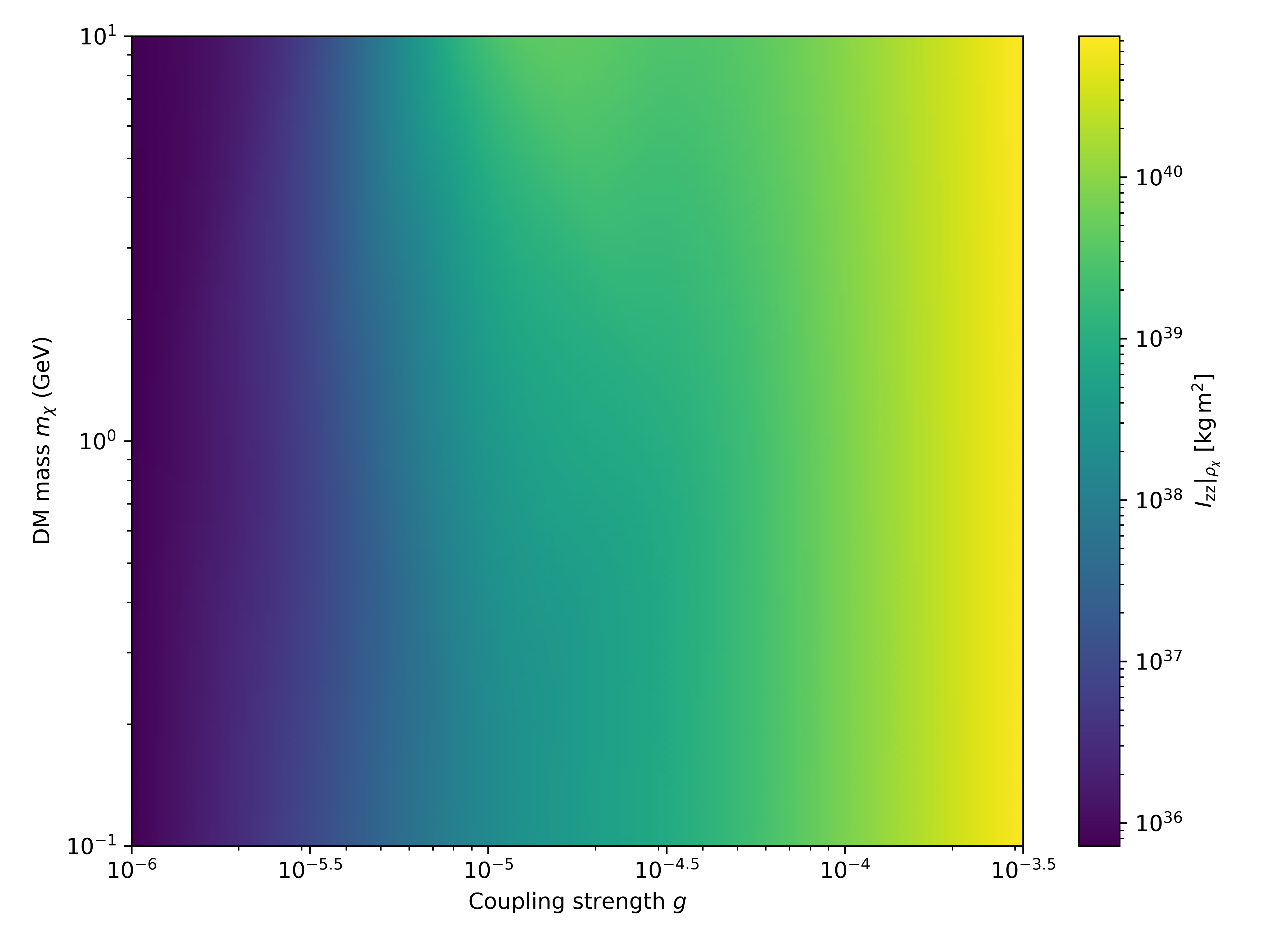}
    \caption{Contribution of DM to the moment of inertia $I_{\rm zz}\big|_{\rho_{\chi}}$ as a function of the DM mass $m_\chi$ and coupling strength $g$. 
    }
    \label{fig:Izz_mchi_g}
\end{figure}

For each $(g, m_\chi)$ pair, $\rho_\chi$ is computed from Eq. \eqref{eq:ferm}, yielding the corresponding DM contribution to DM-admixed neutron stars. Assuming the DM component is distributed uniformly inside the oblate spheroidal neutron star, the principal moment of inertia along the $z$-axis is given by:
\begin{align}
I_{\rm zz}\big|_{\rho_{\chi}}
&= \frac{8\pi}{15} \, a^4 c \, \int_0^{R_{\rm equ}} dr \, \rho_\chi \, r^4
\nonumber \\
& \simeq 1.59 \times 10^{38}\, {\rm kg\,m^2} \left( \frac{\rho_{\chi}}{\rho_{\chi 0}} \right)\,
\label{izzdmns}
\end{align}
where $ \rho_{\chi 0} = 1 \times 10^{18} \, \text{kg / m}^3 $ is the nominal DM density value used throughout this work. Here, we model the neutron star as an oblate spheroid with semi-axes $a=b=1$ and $c = 0.95$, corresponding to a 5\% flattening of the star along the $z$-axis. We take the equatorial and polar radii to be $R_{\textrm{equ}} \simeq a  R_{\rm NS}=10 \,\textrm{km}$ and $R_{\textrm{polar}} \simeq cR_{\rm NS} = 9.5 \, \textrm{km}$.

Therefore, the star's total moment of inertia $I_{\rm zz}$ is
\begin{equation}
    I_{\rm zz} = I_{\rm zz}\big|_{\rho_{\rm B}} + I_{\rm zz}\big|_{\rho_{\chi}}.
    \label{eq:finalizz}
\end{equation}
In Fig.~\ref{fig:Izz_mchi_g}, we show how  $I_{\rm zz}\big|_{\rho_{\chi}}$ depends on $(g, m_{\chi})$.
We see that $I_{\rm zz}\big|_{\rho_{\chi}}$ exhibits a strong and monotonic dependence on $g$ for all values of $m_\chi$. Larger values of $g$ strengthen the repulsive self-interaction in the DM component, which increases its contribution to the total pressure. The resulting change in the mass distribution leads to a systematic increase in $I_{\rm zz}\big|_{\rho_{\chi}}$.

In contrast, the dependence of  $I_{\rm zz}\big|_{\rho_{\chi}}$ on $m_\chi$ at fixed values of $g$ is comparatively weak. 
Within this interval, varying $m_\chi$ primarily rescales the number density of DM particles without significantly altering the overall pressure relative to the baryonic matter component. Consequently, the resulting changes in $I_{\rm zz}\big|_{\rho_{\chi}}$ remain subdominant compared to those induced by varying $g$.

\section{Ellipticity arising from DM anisotropies}
\label{sec:dm_deformation}

The analysis presented above determines the contribution of the DM component to the stellar structure through the energy density $\rho_{\chi}$ (Eq.~\ref{eq:ferm}). However, we assumed that $\rho_\chi$ was constant throughout the star to understand conceptually how $I_{\rm zz}\big|_{\rho_{\chi}}$ varies with $g$ and $m_\chi$. In reality,
the number density $n(r)$ can vary throughout the star, which implies that $k_F$, $x$ and thus $\rho_\chi$ exhibit a radial dependence.

Physically, the radial variation of $n(r)$ reflects the stratified structure of neutron stars, in which this quantity decreases from the inner core to the outer crust. Consequently,  $k_F$ varies from [$2.5$, $10^{-3}$] $\mathrm{fm^{-1}}$ from the core to the surface of the star, inducing a corresponding spatial variation in $\rho_\chi(r)$.

CW emission requires a non-axisymmetric mass distribution in the neutron star. We thus introduce a perturbative deformation of the DM energy density,

\begin{equation}
\rho_\chi(r,\theta,\phi) = \rho_\chi(r)\left[1 + \delta\, f(r)\, Y_{22}(\theta,\phi)\right],
\label{eq:rhochi_perturbed}
\end{equation}
where $Y_{22} (\theta, \phi) \propto \sin^2 \theta \cos(2 \phi)$ is the lowest-order non-axisymmetric quadrupolar mode relevant for CW emission, and $\delta \ll 1$ quantifies the amplitude of an unknown anisotropic distribution. Physically, such deviations from spherical symmetry may arise from rotational deformation, anisotropic stress induced by DM self-interaction, or gravitational coupling between the baryonic and DM components. The function $f(r)$ describes how this anisotropy is radially distributed within the star. It is chosen as an increasing function of radius (e.g., $f(r) \propto r/R$), ensuring that the deformation is more pronounced near the stellar surface.

The perturbation in Eq.~\eqref{eq:rhochi_perturbed} effectively corresponds to a non-axisymmetric density in the DM distribution, analogous to a ``dark mountain'' on the star. Substituting Eq. ~\eqref{eq:rhochi_perturbed} into the moment of inertia tensor

\begin{equation}
I_{ij} = \int d^3x \, \rho_\chi(r,\theta,\phi)\,(r^2\delta_{ij} - x_i x_j),
\end{equation}
we find that the anisotropic contribution appears at the leading order only in the difference between the principal moments

\begin{equation}
I_{\rm xx} - I_{\rm yy} \propto \delta \int_0^R dr \, \rho_\chi(r)\, f(r)\, r^4,
\label{eq:Ixx_minus_Iyy_DM}
\end{equation}
and the leading-order contribution to $I_{\rm zz}$ remains dominated by the isotropic component computed in Eq.~\eqref{izzdmns}. This is a direct consequence of the angular structure of $Y_{22}$, whose integral vanishes when projected onto axisymmetric components.

The resulting ellipticity can therefore be written as,

\begin{align}
\varepsilon  &= \delta \, \frac{\frac{16 \pi a^4 c}{15}\int_0^{R_{\rm equ}} dr \, \rho_\chi(r)\, f(r)\, r^4}{\frac{8 \pi a^4 c}{15}\int_0^{R_{\rm equ}} dr \, \rho_\chi(r) \, r^4}, \nn \\ &= 2 \delta \, \frac{\int_0^{R_{\rm equ}} dr \, \rho_\chi(r)\, f(r)\, r^4}{\int_0^{R_{\rm equ}} dr \, \rho_\chi(r) \, r^4} .
\label{eq:epsilon_DM_final}
\end{align}
The ratio of the integrals in Eq.~\eqref{eq:epsilon_DM_final} depends on the radial profile of $\rho_\chi(r)$ and the specific choice of $f(r)$. 
In our case, we obtain a value of  $\mathcal{O}(1)$ .
As a result, the ellipticity is directly proportional to $\delta$

\begin{equation}
\varepsilon \approx \mathcal{C} \, \delta
\end{equation}
where $\mathcal{C}=5/3$ is the constant \footnote{
For $\rho_\chi = \mathrm{const}$ and $f(r) = r/R$, one finds
\[
\int_0^R \rho_\chi f(r) r^4 dr = \frac{\rho_\chi R^5}{6}, 
\quad
\int_0^R \rho_\chi r^4 dr = \frac{\rho_\chi R^5}{5},
\]
so their ratio is $5/6$, giving $\varepsilon = \frac{5}{3}\delta$. 
Here, $R_{\rm equ} \approx R$ for the oblate spheroid and $\rho_\chi$ is taken to be constant as it is fixed by the DM parameters $(m_\chi, g, m_\phi)$ and represents an effective average DM density within the star.
} determined from Eq.~\eqref{eq:epsilon_DM_final} for the neutron-star density configurations explored in this work.

\begin{figure}[htbp]
\centering
\includegraphics[width=0.8\linewidth]{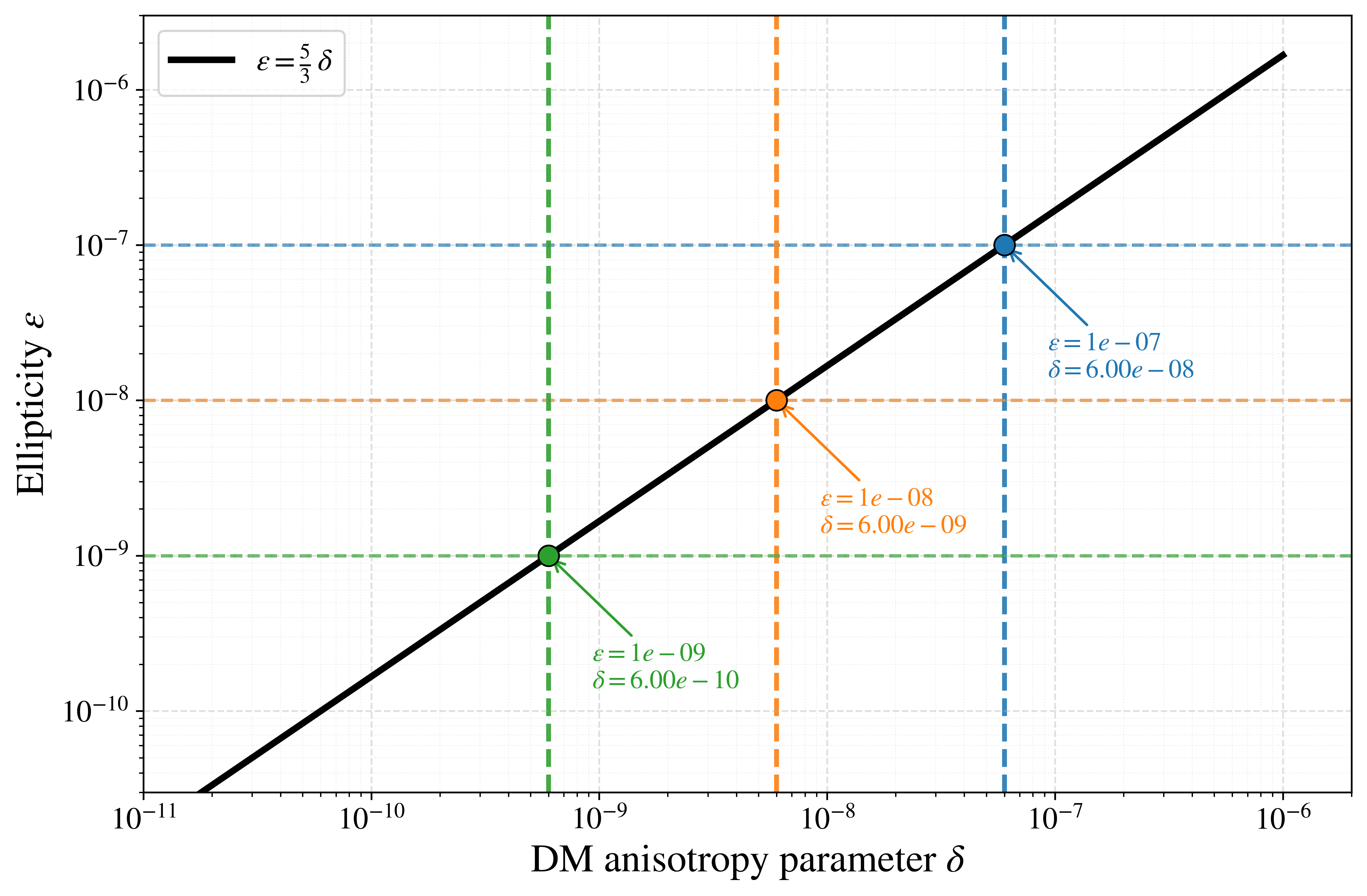}
\caption{
Ellipticity as a function of the DM anisotropy parameter. 
The marked points indicate the ellipticities or anisotropies that we study in this work.
}
\label{fig:epsilon_delta}
\end{figure}

To visualize this relation, Fig.~\ref{fig:epsilon_delta} shows the dependence of $\varepsilon$ on $\delta$. The plot demonstrates a clear linear relation, confirming the analytical expectations $\varepsilon \propto \delta$. The marked points indicate the representative $\varepsilon$ that we will use as benchmark values to constrain the $m_\chi-g$ parameter space later. 

An important implication of Eq.~\eqref{eq:epsilon_DM_final} is that $\varepsilon$ depends weakly on $\rho_\chi(r)$. While $\rho_\chi(r)$ affects both the numerator and denominator integrals, its influence largely cancels in their ratio, leaving $\delta$ as the primary parameter controlling the ellipticity. Consequently, variations in $(m_\chi, g, m_\phi)$ primarily affect the overall normalization of the DM density, but do not significantly alter the linear relation between $\varepsilon$ and $\delta$.

This result provides a clear physical interpretation of the ellipticity values adopted in this work. The ellipticity and DM anisotropy differ by the constant $\mathcal{C}$, indicating that the ellipticity arises naturally from angular perturbations in the DM density in neutron stars.

{ 

Using the computed values of $I_{\rm zz}$ in Fig. \ref{fig:Izz_mchi_g}, we evaluate the corresponding $h_0$ across the ($g, m_{\chi}$)  parameter space. To understand the dependence of $h_0$ on $(g, m_\chi)$ through $I_{\rm zz}$, we fix $\varepsilon=10^{-9}$ ($\delta=0.6\varepsilon$), $d=10$ kpc, and $f_{\rm GW}=111.5$ Hz (the choice of $f_{\rm GW}$ is explained in Sec. \ref{subsec:o3search}), and show the result in Fig. \ref{fig:h0_mchi_fixed_g}.
For fixed $m_{\chi}$, $h_0$ increases steeply with increasing $g$, rising by several orders of magnitude between $g=10^{-6}$ and $g=10^{-3.5}$. The significant scaling is a direct consequence of the high sensitivity of $I_{\rm zz}$ to the coupling strength $g$. Notably, the growth exhibits a non-linear transition around $g \sim 10^{-5}$--$10^{-4.5}$,  where the initial steep increase in $h_0$ becomes less pronounced, producing a visible change in the slope as shown in Fig.~\ref{fig:h0_mchi_fixed_g}. This feature indicates a crossover between two regimes (the core-halo transition happens for those values of $g$ for a DM-admixed neutron star, as discussed in Refs.~\cite {Mahapatra:2024ywx, Liu:2025cwy}), in which the response of $I_{\rm zz}$ to $g$ becomes less efficient. As a result, the rate of increase of $h_0$ with $g$ is temporarily reduced before resuming a smoother monotonic trend at larger couplings.

\begin{figure}[htbp]
    \centering
    \includegraphics[width=0.97\linewidth]{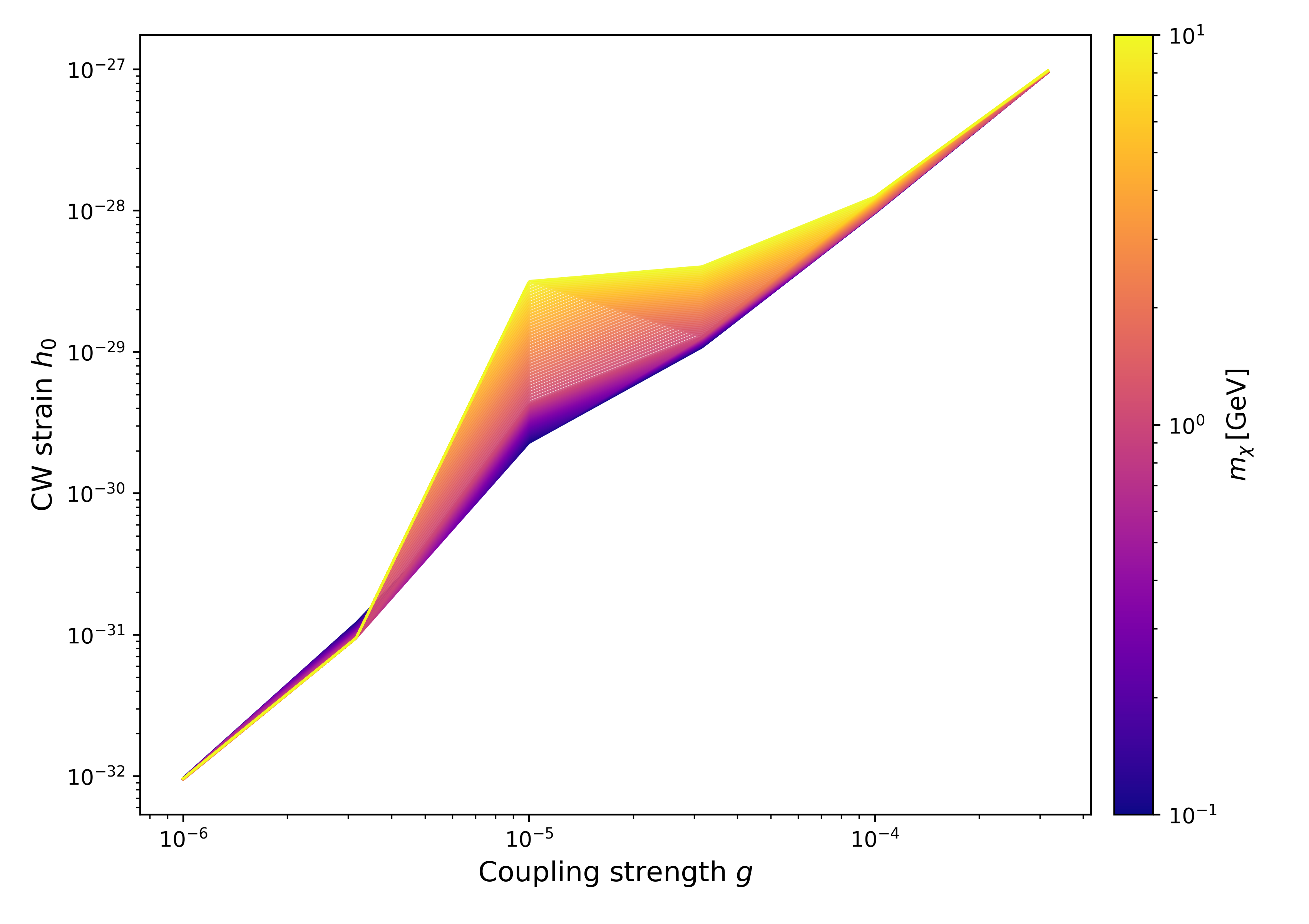}
    \caption{
    CW strain amplitude as a function of coupling strength for $m_\chi \in [0.1,10]$~GeV at $f_{\rm GW}=111.5\,\mathrm{Hz}$, with $\varepsilon=10^{-9}$ ($\delta \sim 10^{-10}$)  and $d=10\,\mathrm{kpc}$. A non-monotonic feature appears at $g \sim 10^{-5}$--$10^{-4.5}$ that corresponds to the core-halo transition.
    }
    \label{fig:h0_mchi_fixed_g}
\end{figure}


Consistent with the weak mass dependence of $I_{\rm zz}$, the variation of $h_0$ with $m_\chi$ at fixed $g$ remains modest across the considered DM mass interval. Overall, Fig.~\ref{fig:Izz_mchi_g} and Fig.~\ref{fig:h0_mchi_fixed_g} demonstrate that $g$ is the primary parameter governing the amplitude of the CW signal, and $m_\chi$ plays a secondary role. 


\section{Methodology}
\label{sec:methodology}

In this section, we explain our methodology to translate upper limits from all-sky searches for CWs to constraints on DM-admixed with particular DM masses and self-interaction coupling strengths. In particular, we take as input to our study upper limits on strain amplitude from an all-sky search of O3 LIGO data~\cite{KAGRA:2022dwb}, in order to constrain DM-admixed neutron stars.

We will first briefly describe the all-sky searches, and then our new method to interpret null results as constraints on DM-admixed neutron stars. 


\subsection{Overview of the O3 all-sky CW search and its upper limits}\label{subsec:o3search}

Using LIGO O3 data, the LVK collaborations and other groups have searched for non-axisymmetric, rotating, isolated neutron stars, but found no evidence of any GW signal. Ref. \cite{KAGRA:2022dwb} analyzed $f_{\rm GW} \in [20,\,2048]~\mathrm{Hz}$ and spin-down values within the range
\begin{equation}
 \dot{f}\in [-10^{-8},+10^{-9}]\,\,  \rm Hz\,s^{-1}
\label{eq:spindo3}
\end{equation}
using multiple pipelines: \textit{Frequency Hough} \cite{Astone_2014}, \textit{Sky Hough} \cite{Krishnan:2004sv, Sintes:2006uc}, \textit{Time domain $\mathcal{F}$- statistic} \cite{Jaranowski:1999pd, LIGOScientific:2014yew}, and \textit{SOAP} \cite{Bayley_2019}. Each pipeline makes different choices regarding noise mitigation and candidate selection, and implements its search independently of the others.
In this work, we adopt the upper limits from the \textit{Frequency Hough} pipeline~\cite{KAGRA:2022dwb}, which analyzed the most complete frequency and spin-down range relative to those of the other pipelines.
Additionally, it achieves the best sensitivity in the frequency band $\sim 100$-$200~\mathrm{Hz}$, with a minimum upper limit of $h_0^{95\%}(f_*) \simeq 1.1 \times 10^{-25}$ at $f_* \simeq 111.5~\mathrm{Hz}$.

Within the context of this analysis, our methodology involves three steps:

\begin{enumerate}
    \item Verify whether the predicted spin-down from DM-admixed neutron stars lies within the spin-down range given in Eq. \eqref{eq:spindo3}.
    \item Cast the \textit{Frequency-Hough} $95\%$ upper limits $h_0^{95\%}$ into the maximum detectable ellipticities on DM-admixed neutron stars for different $g,m_\chi$.
    \item Identify regions of the parameter space ($g, m_\chi$) that would have produced a strain greater than the upper limits, which excludes the existence of DM-admixed neutron stars with those DM masses and self-interaction couplings.
\end{enumerate}

\begin{figure}[htbp]
    \centering
    \includegraphics[width=0.97\linewidth]{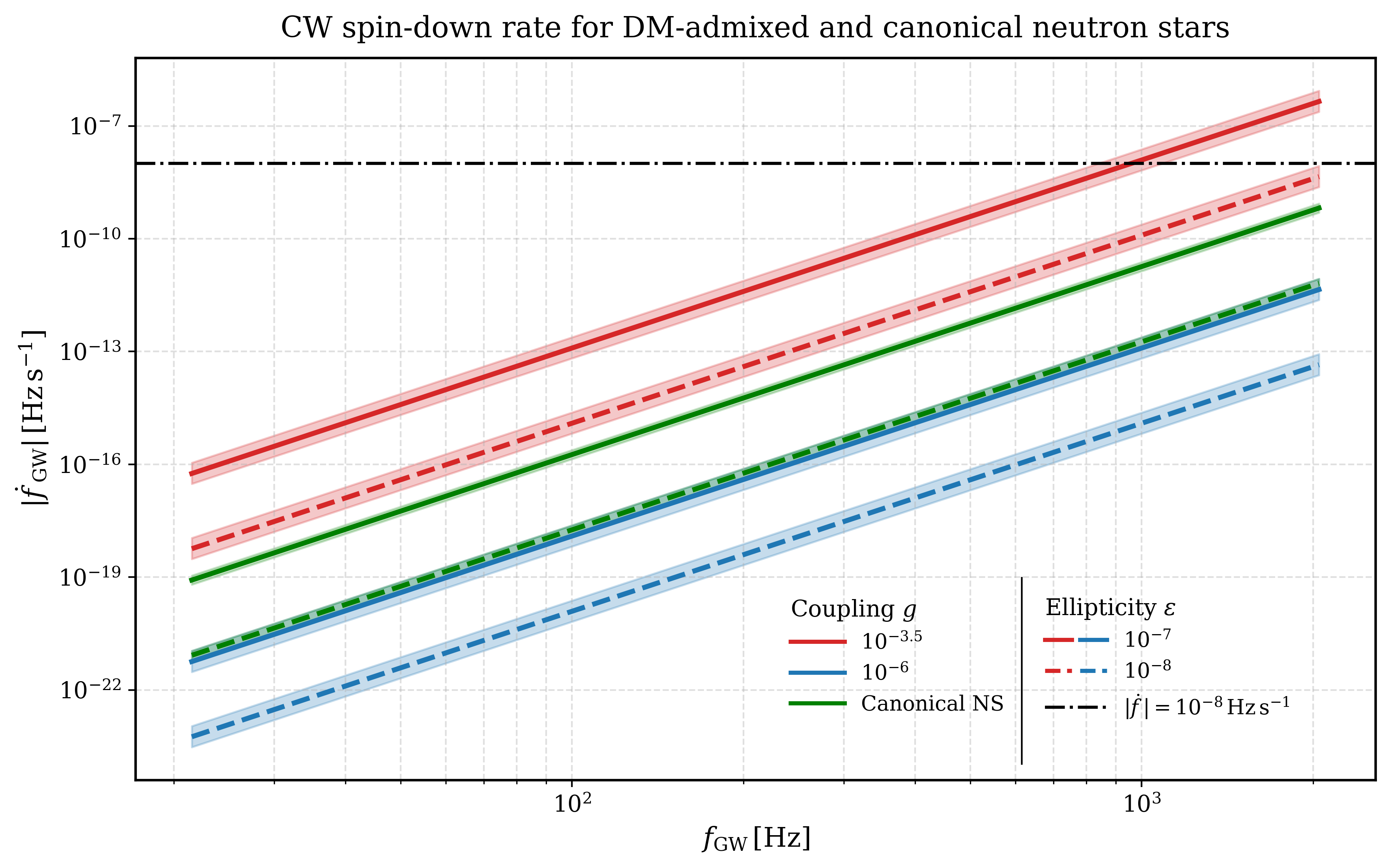}
    \caption{
    Spin-down $|\dot f_{\rm GW}|$ for DM-admixed neutron stars across the  frequency range analyzed in the LVK O3 CW search. We plot $|\dot f_{\rm GW}|$ for two representative couplings, $g=10^{-3.5}$ (pink) and $10^{-6}$ (blue), for two ellipticities $\varepsilon = 10^{-8}$ (dashed; $\delta \sim 10^{-9}$) and $10^{-7}$ (solid; $\delta \sim 10^{-8}$), and for an ordinary neutron star (green). The horizontal dashed line indicates the maximum spin-down value analyzed in the search.
    Only the configuration $(g=10^{-3.5},\,\varepsilon=10^{-7})$ extends above the dashed line for $f_{\rm GW}\gtrsim1000~\mathrm{Hz}$.
    }
    \label{fig:fdot_DM_vs_canonical}
\end{figure}

\subsection{DM-admixed neutron-star spin-downs within the searched parameter space}
\label{subsec:spindown}

The spin-down rate of the neutron star depends on $\varepsilon$ and $I_{\rm zz}$ (Eq. \eqref{eq:fdot}). Since the O3 search is restricted to the spin-down range defined by Eq.~\eqref{eq:spindo3}, we must verify that the predicted spin-downs from DM-admixed neutron stars fall within this range.

{ We evaluate $\dot f_{\mathrm{GW}}$ as a function of $f_{\mathrm{GW}}$ corresponding to different $\varepsilon$ and $(g, m_\chi)$ values in the parameter space. In Fig.~\ref{fig:fdot_DM_vs_canonical}, we show the variation of $|\dot f_{\mathrm{GW}}|$ with $f_{\mathrm{GW}}$ for the representative couplings $g = 10^{-3.5}$ and $g = 10^{-6}$ for $\varepsilon=10^{-8},10^{-7}$, and compare those curves with the behavior that is expected for canonical neutron stars.
\begin{figure*}[htbp]
    \centering
    \includegraphics[width=0.97\linewidth]{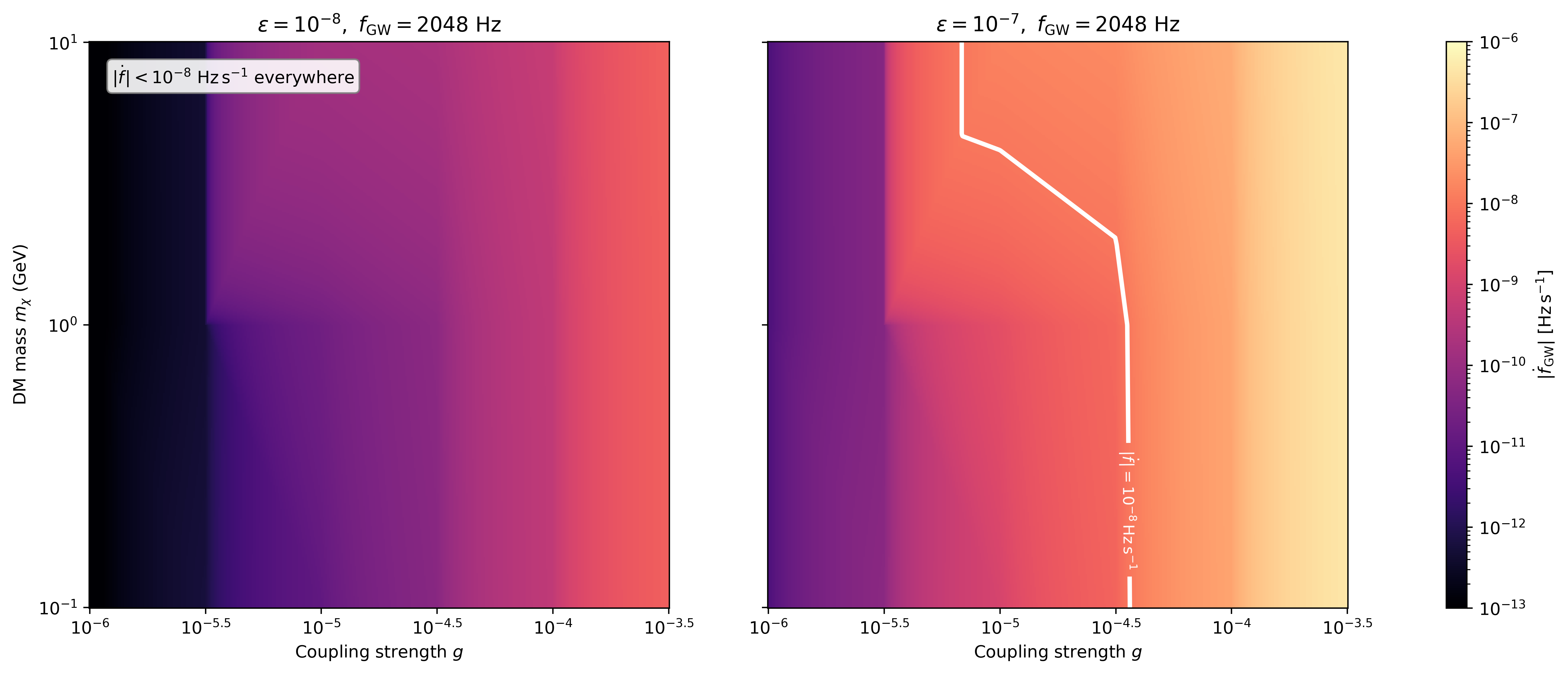}
    \caption{
    Spin-down $|\dot f_{\rm GW}|$ evaluated over the $(g, m_\chi)$ parameter space at $f_{\rm GW}=2048~\mathrm{Hz}$. The white contour indicates the maximum spin-down analyzed in the O3 all-sky search. Left panel: $\varepsilon = 10^{-8}$ ($\delta \sim 10^{-9}$), all spin-downs lie within the searched range. Right panel: $\varepsilon = 10^{-7}$ ($\delta \sim 10^{-8}$), spin-downs to the right of the contour lie outside the searched range. 
    }
    \label{fig:fdot_exclusion}
\end{figure*}
For moderate deformations, i.e. $\varepsilon  = 10^{-8}$, the predicted spin-downs remain within the range given in Eq.~\eqref{eq:spindo3} across the full frequency range for the chosen couplings. However, when $\varepsilon= 10^{-7}$ and $g =10^{-3.5}$, the predicted spin-downs exceed the {maximum} given in Eq.~\eqref{eq:spindo3} at higher frequencies. This implies that the O3 search is insensitive to quickly rotating, highly deformed DM-admixed neutron stars.

In Fig.~\ref{fig:fdot_exclusion}, we show how $\dot{f}_{\rm GW}$ changes across the $(g,m_\chi)$ parameter space at $f_{\rm GW}=2048~\mathrm{Hz}$, the maximum frequency analyzed in O3, in order to obtain a conservative understanding of the kinds of DM-admixed neutron stars to which we could be sensitive. 
The white line delineates the region of parameter space for which the predicted spin-down satisfies the condition given in Eq. \eqref{eq:spindo3}. For $\varepsilon = 10^{-8}$, all points in the $(g,m_\chi)$ parameter space yield spin-down values within this range; however, for $\varepsilon = 10^{-7}$, a subset of the parameter space at larger $g$ produces spin-down exceeding the maximum searched in O3. Thus, CW analyses are sensitive to a wide range of deformed, DM-admixed neutron stars, even at the highest frequency analyzed.

\begin{figure}[htbp]
    \centering
    \includegraphics[width=0.97\linewidth]{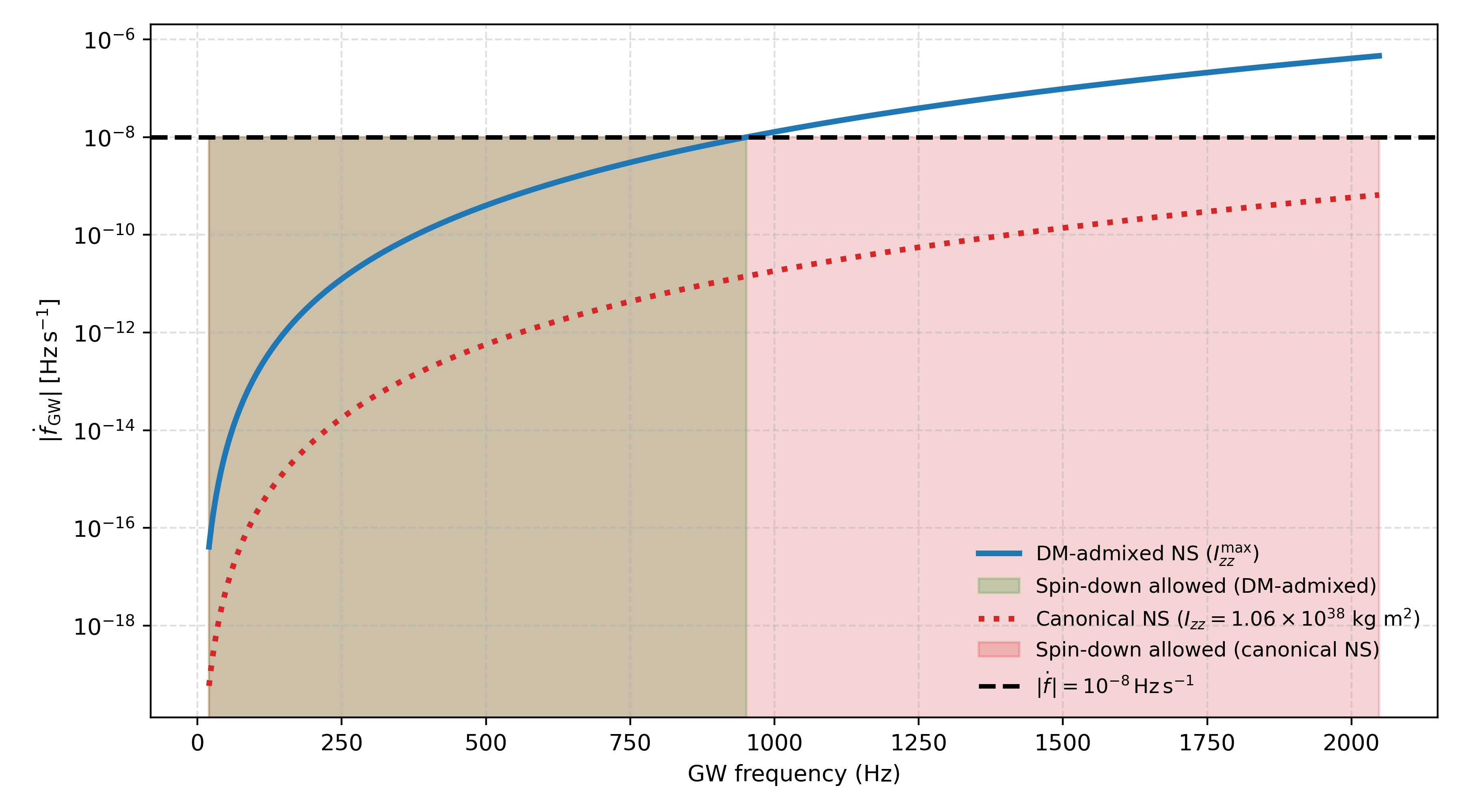}
    \caption{
    Spin-down $|\dot f_{\rm{GW}}|$ as a function of GW frequency for $\varepsilon = 10^{-7}$ ($\delta \sim 10^{-8}$). The solid blue curve shows the maximum $|\dot f_{\rm GW}|$ calculated using the largest $g,m_\chi$ and $\varepsilon$ considered here, while the red dotted curve corresponds to a canonical neutron star without DM. The horizontal dashed line denotes the maximum spin-down value analyzed in the O3 all-sky search. The shaded green (pink) region highlights the frequency range for which all DM-admixed stars (canonical neutron stars) lie within the spin-down range covered by the O3 search.
    }
    \label{fig:spin_down_band}
\end{figure}

To further illustrate how $\dot f_{\rm GW}$ changes with frequency, we show in Fig.~\ref{fig:spin_down_band} the maximum spin-down obtained over the $(g, m_\chi)$ parameter space as a function of $f_{\rm GW}$ for $\varepsilon = 10^{-7}$, the largest ellipticity we consider in this work that would thus lead to the largest $\dot{f}_{\rm GW}$. 
The solid blue curve represents the maximum spin-down over the $(g,m_\chi)$ parameter space for DM-admixed neutron stars, while the red dotted curve corresponds to the spin-down of a canonical neutron star without DM. The shaded green region identifies the frequency ranges in which all DM-admixed neutron-star models yield spin-downs within the O3 search range, while the shaded pink regions show the corresponding ranges for canonical neutron stars. At fixed $\varepsilon$ and $f_{\rm GW}$, the larger $I_{\rm zz}$ in DM-admixed neutron stars leads to a bigger spin-down than for ordinary neutron stars. Consequently, the blue curve exceeds the bound in Eq.~\eqref{eq:spindo3} at sufficiently high frequencies, whereas the canonical neutron star prediction remains below this value across the full frequency range.

While Fig.~\ref{fig:fdot_DM_vs_canonical} and Fig.~\ref{fig:fdot_exclusion} show $|\dot{f}_{\rm GW}|$ for representative $(g,m_\chi)$ at a fixed frequency, Fig.~\ref{fig:spin_down_band} identifies the frequency at which the maximal spin-down crosses the bound in Eq.~\eqref{eq:spindo3}, providing complementary ways of understanding how all-sky searches can be sensitive to DM-admixed neutron stars.


\begin{figure*}[htbp]
\centering
\includegraphics[width=0.97\textwidth]{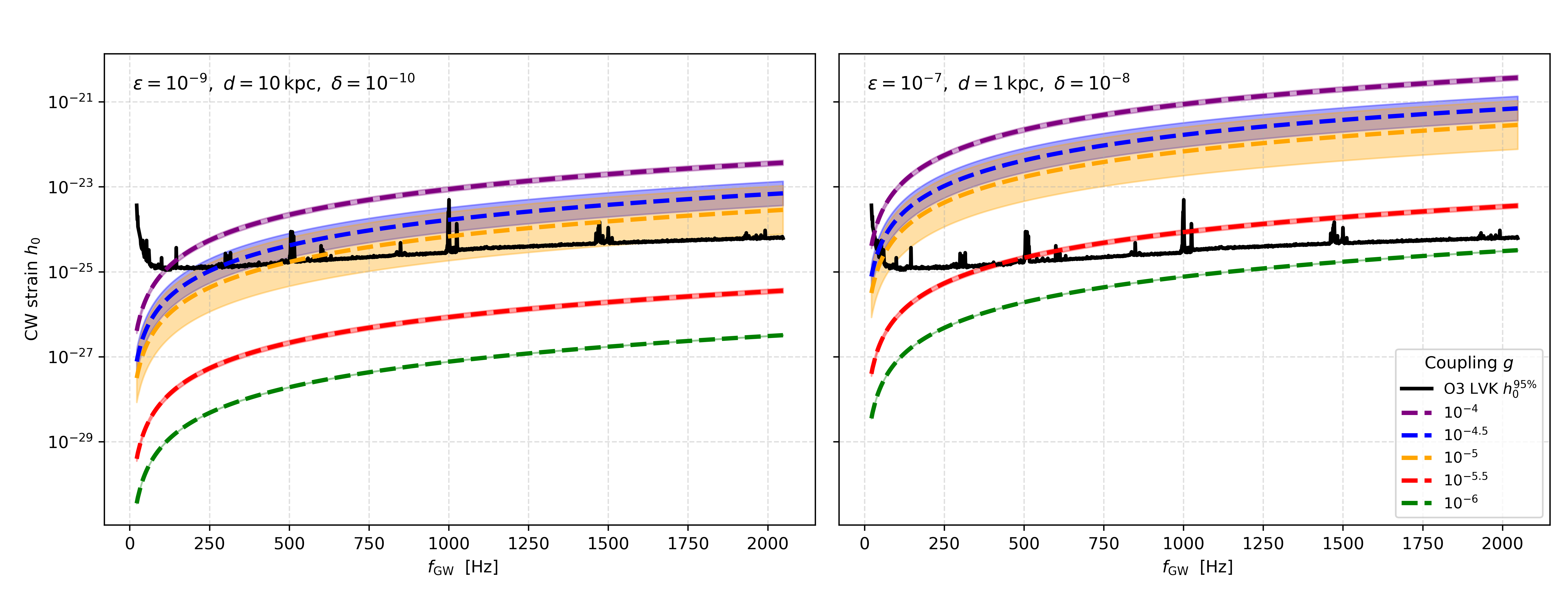}
\caption{Strain amplitude for DM-admixed neutron stars compared with the LVK O3 $h_0^{95\%}$ upper limits across the $g,m_\chi$ parameter space. Left: $\varepsilon = 10^{-9}$ ($\delta \sim 10^{-10}$), $d = 10~\mathrm{kpc}$ (minimum ellipticity, maximum distance). Right: $\varepsilon = 10^{-7}$ ($\delta \sim 10^{-8}$), $d = 1~\mathrm{kpc}$ (maximum ellipticity, minimum distance). Colored dashed curves denote different coupling strengths $g$, and shaded regions indicate variation over $m_\chi \in [0.1,10]~\mathrm{GeV}$.
}
\label{fig:h0_combined_extremes}
\end{figure*}

\subsection{Constraining the ($g, m_\chi$) parameter space with O3 strain upper limits}\label{subsec:constrgmchi}

Having established which DM-admixed neutron stars lie within the spin-down range covered by the O3 search, we now compute the expected strain amplitudes of such signals as a function of mass and coupling strength using Eq. \eqref{eq:h0} and Eq. \eqref{izzdmns}. We then compare these amplitudes to upper limits derived from the O3 search to identify portions of the $m_\chi-g$ parameter space that would have produced detectable CW signals in O3, i.e. $ h_0 > h_0^{95\%}$, and can thus be excluded.

Though the upper limits are given between $[20,2048]$ Hz, we pick the frequency at which the smallest strain amplitude is derived, as mentioned in Sec. \ref{subsec:o3search}. Since our goal is to constrain the DM space by finding \emph{any} DM-admixed neutron star, not the population of DM-admixed neutron stars, we are free to choose the best upper-limit value. 

\section{Constraints on dark-matter admixed neutron stars}
\label{sec:results}



\subsection{Comparison of strain amplitudes with O3 upper limits}
\label{subsec:h0_bands}

We first examine the CW strain amplitude using Eq.~\eqref{eq:h0} and $I_{\rm zz}(g,m_\chi)$ obtained from Eq.~\eqref{eq:finalizz} and shown in Fig.~\ref{fig:Izz_mchi_g}. The resulting strain is compared with the O3 upper limits following the procedure in Sec.~\ref{sec:methodology}.

Fig.~\ref{fig:h0_combined_extremes} shows the CW strain as a function of $f_{\rm GW}$ for two representative choices of ellipticity and source distance. 
The left panel corresponds to the most conservative scenario, i.e. assuming the minimum ellipticity ($\varepsilon = 10^{-9}$, $\delta \sim 10^{-10}$) and maximum distance ($d = 10~\mathrm{kpc}$), while the right panel shows the most optimistic case, i.e. assuming the maximum ellipticity ($\varepsilon = 10^{-7}$, $\delta \sim 10^{-8}$) and minimum distance ($d = 1~\mathrm{kpc}$). Together, these panels bracket the weakest and strongest CW signals we consider here.

In the left panel, the predicted strain exceeds the O3 upper limits for $g > 10^{-4}$, while for smaller couplings, the strain remains below the observational bound and thus no DM parameters can be excluded.


\begin{figure*}
    \centering
    \includegraphics[width=0.97\textwidth]{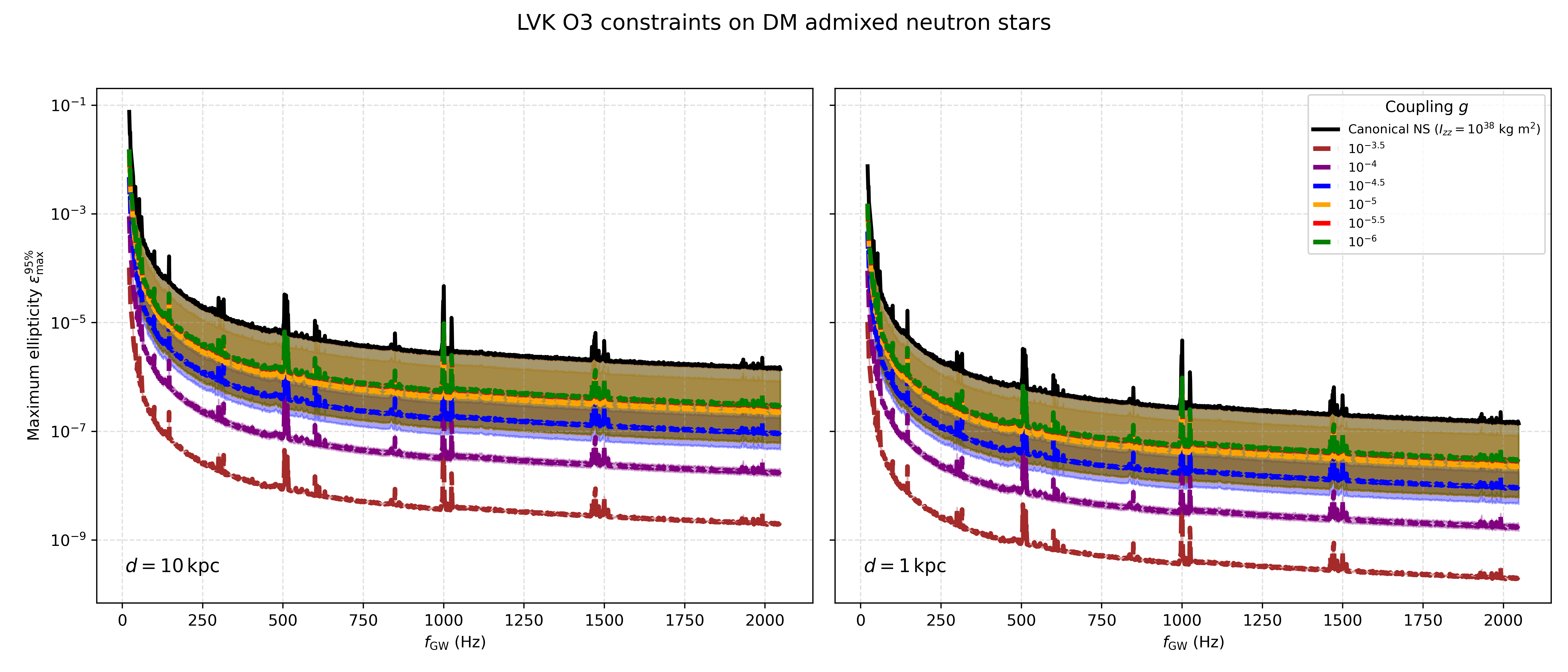}
    \caption{Constraints on the maximum sustainable ellipticity $\varepsilon_{\max}^{95\%}$ of DM-admixed neutron stars as a function of frequency derived from the O3 all-sky search upper limits. Colored curves show the allowed range over $m_{\chi} \in [0.1,10]~\mathrm{GeV}$ for different couplings, while the solid black curve denotes the corresponding ellipticity bound for a canonical neutron star without DM. Left panel: $d=10~\mathrm{kpc}$; right panel: $d=1~\mathrm{kpc}$.}
    \label{fig:ellipticty_constraint}
\end{figure*}

In the right panel, the strain is enhanced due to the combined effects of larger ellipticity and smaller distance. In this case, all couplings down to $g = 10^{-5.5}$ are excluded, and only very weak self-interactions, $g \lesssim 10^{-6}$, remain compatible with the upper limits.

A notable feature in both panels is the emergence of a band-like structure for intermediate couplings $g \in [10^{-5} - 10^{-4}]$, where the strain curves partially overlap. We find that $h_0$ exhibits a mild non-monotonic dependence on $g$ in this regime, previously observed in Fig.~\ref{fig:h0_mchi_fixed_g} in Sec.~\ref{subsec:dmns}.

\subsection{Constraints on dark mountains}

We can also convert the O3 strain upper limits $h_0^{95\%}$ into frequency-dependent upper bounds on the ellipticity $\varepsilon^{95\%}_{\rm max}$ using Eq. \eqref{eq:h0}. In  Fig.~\ref{fig:ellipticty_constraint}, we show $\varepsilon^{95\%}_{\rm max}$ as a function of $f_{\rm GW}$ for DM-admixed neutron stars with different values for $g$ (colored curves) and for a canonical neutron star with moment of inertia $I_{0}$ (solid black curve), assuming source distances of $d=10~\mathrm{kpc}$ (left panel) and $d=1~\mathrm{kpc}$ (right panel). 

A key feature of Fig.~\ref{fig:ellipticty_constraint} is that the presence of DM always increases $I_{\rm zz}$, 
which lowers the value of $\varepsilon^{95\%}_{\rm max}$ for a given $h_0^{95 \%}$. 
These results demonstrate that the enhancement of $I_{zz}$ induced by DM can tighten the ellipticity constraints by up to several orders of magnitude relative to the canonical neutron star scenario.

Expressing the results in terms of $\varepsilon^{95\%}_{\rm max}$ is essential as it directly quantifies the largest quadrupolar deformation compatible with current CW non-detections for DM-admixed neutron stars. In particular, the dependence of $\varepsilon^{95\%}_{\rm max}$ on $g$ provides a direct link between observational upper limits and the strength of DM self-interactions, complementing the exclusion regions in the $(g, m_\chi)$ parameter space discussed below.

\begin{figure*}[htbp]
\centering
\includegraphics[width=0.99\textwidth]{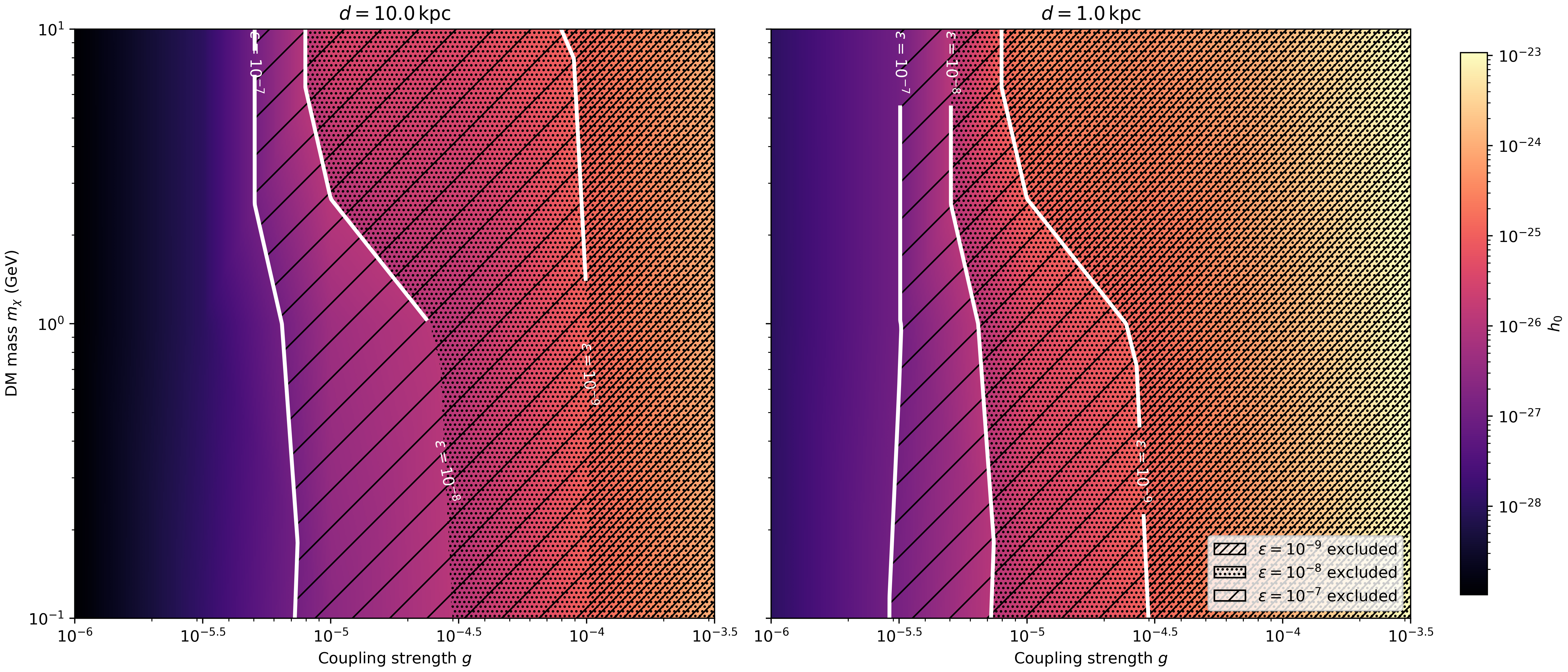}
\caption{Exclusion regions in the $(g, m_\chi)$ plane for ellipticities $\varepsilon=10^{-9},\,10^{-8},\,10^{-7}$ (corresponding to DM anisotropies $\delta \sim 10^{-10},10^{-9},10^{-8}$, respectively), for distances of $d=10~\mathrm{kpc}$ (left) and $d=1~\mathrm{kpc}$ (right) at a GW frequency of $f_*=111.5$ Hz. The color indicates the predicted GW strain amplitude $h_0$ for different values of $g$ and $m_\chi$. White contours denote when the predicted GW amplitude equals the best upper limit from O3, $h_0^{95\%}(f_*)=1.1\times 10^{-25}$, separating allowed and excluded regions. The excluded regions are indicated by hatch patterns corresponding to each ellipticity. }
\label{fig:exclusion_contours_combined}
\end{figure*}

\subsection{Exclusion contours in the $(g, m_\chi)$ plane}
\label{subsec:exclusion_contours}

We now translate the strains shown in Fig.~\ref{fig:h0_combined_extremes} into constraints on the $(g, m_\chi)$ parameter space, following the procedure outlined in Sec.~\ref{sec:methodology}.

Fig.~\ref{fig:exclusion_contours_combined} displays the resulting exclusion maps for three chosen ellipticities $\varepsilon = 10^{-9},\,10^{-8},\,10^{-7}$ ($\delta \sim 10^{-10},10^{-9},10^{-8}$, respectively) for $d = 10~\mathrm{kpc}$ (left column) and $d = 1~\mathrm{kpc}$ (right column). The colormap represents the GW strain, while the white contours indicate when the calculated strain amplitude equals the upper-limit value. Observationally excluded regions are distinguished by hatch patterns for each $\varepsilon$.


The exclusion boundaries depend strongly on the coupling $g$ but show only mild sensitivity to $m_\chi$. This behavior follows from the dependence of $I_{\rm zz}$ on $g$ and its impact on $h_0$, as shown in Fig.~\ref{fig:Izz_mchi_g} and Fig.~\ref{fig:h0_mchi_fixed_g}, respectively. Additionally, reducing the source distance shifts the exclusion boundary towards smaller couplings. 

In the left panel (for farther sources), at $\varepsilon = 10^{-9}$, the strain remains below the O3 limits across most of the parameter space, and we only exclude $g^{95\%} \gtrsim 10^{-4}$. At $\varepsilon = 10^{-8}$, the contours remain confined to the intermediate and strong couplings and we can only exclude $g^{95\%} \gtrsim [10^{-5},10^{-4.5}]$ for larger and smaller $m_\chi$, respectively. Even for $\varepsilon = 10^{-7}$, the exclusion boundary only reaches $g^{95\%} \gtrsim 10^{-5.25}$.

In the right panel (for closer sources), the overall structure of the exclusion contours remains similar, but the boundaries are shifted towards larger values of $g$. 
For $\varepsilon = 10^{-7}$, however, the exclusion contour penetrates deep into the weak-coupling regime, ruling out $g^{95\%} \gtrsim 10^{-5.5}$ over a broad range of $m_\chi$.

Our results show that the source distance controls the overall position of the exclusion boundary, while the chosen ellipticity determines how far the exclusion region extends into weak-coupling regime. Larger ellipticities enhance the strain amplitude, thereby extending the region of parameter space that can be probed by CW searches. This highlights the importance of highly deformed neutron stars as potential probes of DM-induced modifications to stellar structure.


\section{Implications for CW searches next-generation detectors}
\label{sec:implications}



Searching for DM-admixed neutron stars and setting constraints at galactic distances ($d \sim 10~\mathrm{kpc}$) is important because it corresponds to where we expect a large number of neutron stars to be in the Milky Way. However, based on Fig.~\ref{fig:exclusion_contours_combined}, we can only probe the smallest couplings in our parameter space for larger ellipticities at distances of 1 kpc. This suggests that improvements in detector sensitivity could extend the search to a substantially larger population of DM-admixed neutron stars throughout the Galaxy.

Fortunately, next-generation detectors such as CE \cite{Reitze:2019iox, Evans:2023euw, DiGiovanni:2025rhy} and ET \cite{ET:2019dnz, Branchesi:2023mws, ET:2025xjr} are expected to enhance CW sensitivity by more than an order of magnitude, as shown in Fig.~\ref{fig:CE_ET_exclusion_plot} for GW frequencies of $f_* = 133.49~\mathrm{Hz}$ and $f_* = 271.73~\mathrm{Hz}$, and minimum strain amplitudes of $h_0^{95\%} = 4.55 \times 10^{-27}$ and $6.88 \times 10^{-27}$, for CE and ET, respectively. The improved strain sensitivity shifts the white contours towards the weak-coupling regime $g \gtrsim 10^{-5.5}$ across the DM mass range considered here. The excluded regions to the right of the white contours are indicated by hatch patterns for each ellipticity. The projected excluded regions greatly improve upon those that we derived from O3.


Overall, these results demonstrate that CW searches offer a way to constrain DM self-interactions through their impact on neutron stars, and improvements in detector sensitivity will significantly extend this reach.

 \begin{figure*}[htbp]
    \centering
    \includegraphics[width=0.99\textwidth]{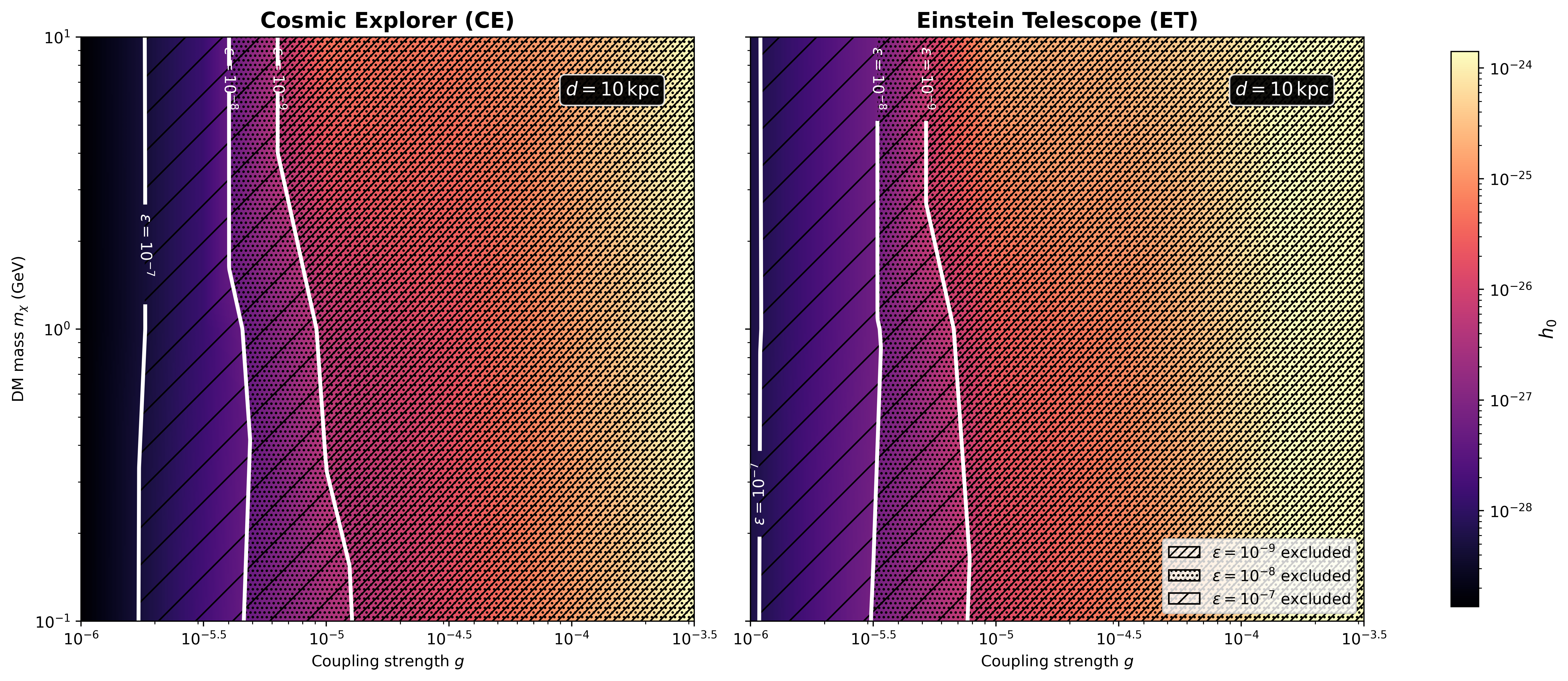}
    \caption{Same as in Fig. \ref{fig:exclusion_contours_combined}, but the plot represents projected exclusion regions using the expected sensitivity of CE (left) and ET (right) at $d=10$ kpc. The projections are computed at GW frequencies $f_* = 133.49~\mathrm{Hz}$ and $f_* = 271.73~\mathrm{Hz}$ with corresponding minimum strain amplitude of $h_0^{95\%} = 4.55 \times 10^{-27}$ and $h_0^{95\%}=6.88 \times 10^{-27}$ for CE and ET, respectively. 
    }
    \label{fig:CE_ET_exclusion_plot}
\end{figure*}

}

\section{Conclusion}
\label{sec:conclusion}

In this work, we have investigated CW emission from neutron stars composed of self-interacting fermionic DM.
Within a two-fluid model based on relativistic mean-field baryonic matter coupled to a Yukawa-type DM interaction, we showed that DM can induce non-zero ellipticities in isolated neutron stars if it is distributed anisotropically inside the star, and modify the moment of inertia with respect to stars comprised of ordinary matter. We have thus established a link between DM astrophysics and CWs. 

We also demonstrated that larger moments of inertia in DM-admixed neutron stars lead to stronger constraints on their maximum ellipticities than from ordinary stars.
Thus, we have, for the first time, set constraints on the ``dark mountains'' sustained by DM-admixed neutron stars.

By using upper limits from an all-sky search for CWs in LIGO O3 data, we have excluded the existence of DM-admixed neutron stars in certain portions of the parameter space for assumed distances and ellipticities of DM-admixed neutron stars. In particular, we rule out couplings as small as $10^{-5.5}$ ($10^{-4}$) in the most optimistic (pessimistic) assumptions for distance and ellipticity. Our analysis thus demonstrates sensitivity to coupling strengths that are over two orders of magnitude smaller than previous works \cite{Guha:2024pnn, Mariani:2023wtv}, although we consider a different DM mass range. Moreover, in contrast to astrophysical observations of $\sigma/m_\chi $ that probe large scales, we provide a way to constrain DM on the scale of individual astrophysical objects.

The future for such analyses of DM-admixed neutron stars is bright. As ground-based detectors continue to improve in sensitivity, and future ones come online, we demonstrate that CW searches can be sensitive to even weaker self-interaction couplings, motivating the possibility of actually being able to detect DM-admixed neutron stars, or further constrain their existence. In this sense, future CW observations have the potential to transform rotating neutron stars into powerful probes of DM physics, providing a complementary avenue for testing self-interactions in regimes that are inaccessible to laboratory experiments or telescopes.

\appendix

\section{Self-Interaction Cross-Section}
{ 
We compute the cross-section / DM mass ratio for representative parameters that we consider in this work, and show that it is compatible with the bounds inferred from the Bullet cluster. The DM-DM self-interaction cross-section for $\chi \bar{\chi} \to \chi \bar{\chi}$ mediated by \(\phi\) mediator and coupling $g$ is:
\begin{align}
   &\sigma = \frac{g^{4} m_{\chi}^{2}}{4 \pi m_{\phi}^4 } \nn \\
   &\simeq 3.1\times 10^{-25} ~\mathrm{cm^{2}}\left(\frac{g}{10^{-5}}\right)^4\left(\frac{1~\mathrm{keV}}{m_\phi}\right)^4 \left(\frac{m_\chi}{1~\mathrm{GeV}}\right)^2,
\end{align}
which, for the reference values in the above equation, implies $\frac{\sigma}{m_{\chi}}\simeq 0.1~\mathrm{cm^{2}/g}$.




\section{Mean free path constraint for self-interacting DM inside a neutron star}

{For DM to remain confined inside the neutron star, its mean free path should satisfy} 
\begin{align}
\lambda &= \frac{1}{n_{\chi}\,\sigma} \nn \\
&\simeq2\times 10^{-16}~\mathrm{m} \left(\frac{3.1\times 10^{-25}\mathrm{cm^2}}{\sigma}\right)  \ll R_{\rm NS}
\label{eq:mfp_condition}
\end{align}
where $n_{\chi} = 10^{-3} n_0$ is the DM number density, $n_0 = 0.16~ \rm{fm^{-3}}$ is the nuclear matter number density~\cite{Guha:2024pnn, Iqbal:2026kjj}, and $\sigma$ is the DM self-interaction cross-section.

We can now compute $\lambda$ by varying $m_\chi$ and $g$ within our analyzed parameter space, and find that $\lambda$ is between 14 and 20 orders of magnitude lower than the radius of the neutron star, indicating that self-interacting DM is always trapped inside the neutron-star core.

}

\section*{Acknowledgements}
This material is based upon work supported by NSF's LIGO Laboratory, which is a
major facility fully funded by the National Science Foundation. The authors gratefully thank David Ian Jones for his comments regarding the origin of the dark mountains. P.M would like to thank BITS Pilani, K K Birla Goa campus for the fellowship support; Subhadip Sau and Apratim Ganguly for related discussion on two-fluid DM-admixed neutron stars models; Ayush Hazarika for discussions on theory related to continuous gravitational waves. A.L.M. is supported by the National Natural Science Foundation of China (NSFC) under Grants No. 12347103 and 12547104, and by the Fundamental Research Funds for the Central Universities. {PKD gratefully acknowledges the financial support of ANRF Grant No. CRG/2023/008877.}

\bibliographystyle{./utphys1}
\bibliography{references}

\end{document}